\documentclass[letterpaper,twocolumn,10pt]{article}

\usepackage{graphicx}             
\usepackage[caption=false]{subfig}
\usepackage{url}                  
\usepackage[T1]{fontenc}
\usepackage{nicefrac}             
\usepackage{indentfirst}          
\usepackage[super, negative]{nth}
\usepackage{amsfonts}
\usepackage{amssymb}
\usepackage{multirow}
\usepackage{balance}
\usepackage{subfloat}

\usepackage{xspace}
\usepackage{array,booktabs}
\usepackage{latexsym}
\usepackage{pifont}
\usepackage{colortbl}

\usepackage{algorithm}
\usepackage[noend]{algpseudocode}
\usepackage{listings}

\usepackage{xcolor}

\newcommand{\dfn}[1]{\textit{#1}}            
\newcommand{\ioam}{\textsc{Ioam}\xspace}
\newcommand{\traceroute}{\texttt{traceroute}\xspace}

\makeatletter
\newcommand\footnoteref[1]{\protected@xdef\@thefnmark{\ref{#1}}\@footnotemark}
\makeatother

\graphicspath{{.}{./Pictures/}}

\definecolor{comment}{RGB}{2,112,10}
\begin{document}

\title{Implementation of \ioam for IPv6 in the Linux Kernel}
\author{Justin Iurman{$^{\ast}$}, Benoit Donnet{$^{\ast}$}, Frank Brockners{$^{\ddag}$}\\
$\ast$ Universit\'e de Li\`ege, Montefiore Institute -- Belgium\\
$\ddag$ Cisco
}
\date{}

\maketitle

\begin{abstract}
In-situ Operations, Administration, and Maintenance (\ioam) is currently under
standardization at the IETF.  It allows for collecting telemetry and operational 
information along a path, within the data packet, as part of an existing 
(possibly additional) header.  This paper discusses the very first 
implementation of \ioam for the Linux kernel with IPv6 as encapsulation
protocol. We also propose a first preliminary evaluation of our implementation
under a controlled environment. Our \ioam implementation is available as open source.
\end{abstract}

\section{Introduction}\label{intro}
\dfn{Operations, Administration, and Maintenance} (OAM) refers to a set of
techniques and mechanisms for performing fault detection and isolation and for
performance measurements.  Multiple OAM tools have been developed, throughout
the years, for various layers in the protocol stack~\cite{rfc7276}, going from 
basic \traceroute~\cite{traceroute} to Bidirectional Forwarding Detection
(BFD~\cite{rfc5880}).  Recently, OAM has been pushed further through
\dfn{In-Situ OAM} (\ioam)~\cite{ioam-requirements,ioam-draft}.  The term
``In-Situ'' directly refers to the fact that the OAM and telemetry data is 
carried within data packets rather than being sent within packets specifically
dedicated to OAM. The \ioam traffic is embedded in data traffic, but not part of 
the packet payload.

The well-known IPv4 Record-Route option~\cite{rfc791} can be considered as an
\ioam mechanism.  However, compared to the Record-Route option, \ioam comes with
multiple advantages: ($i$) \ioam is not limited to 40 bytes as for the
Record-Route option (recording so a maximum of 9 IPv4 addresses) ($ii$) \ioam
allows different types of information to be captured including not only path
tracing information but additional operational and telemetry information such as
timestamps, sequence numbers, or even generic data such as queue size,
geo-location of the node that forwarded the packet; ($iii$) \ioam allows one to
record the path taken by a packet within a fixed amount of added data; ($iv$)
\ioam data fields are defined in a generic way so that they are independent from
the protocol that carries them; ($v$) finally, \ioam offers the ability to
actively process information in the packet.  For instance, \ioam allows one to
prove in a cryptographically secure way that a packet really followed a
pre-defined path using a traffic steering method, such as service chaining
(e.g., NFV service chaining) or traffic engineering (e.g., through Segment
Routing~\cite{rfc8402}).

In a nutshell, \ioam gathers telemetry and operational information along a path,
within the data packet, as part of an existing (possibly additional) header.  It
is encapsulated in IPv6 packets as an IPv6 HopByHop extension
header~\cite{hbh_header}.  Typically, \ioam is deployed in a given domain,
between the Ingress and the Egress or between selected devices within the
domain.  Each node involved in \ioam may insert, remove, or update the extension
header. \ioam data is added to a packet upon entering the domain and is removed
from the packet when exiting the domain. There exist four \ioam types for which
different \ioam data fields are defined. ($i$) the \dfn{Pre-allocated Trace
Option}, where space for \ioam data is pre-allocated; ($ii$) the
\dfn{Incremental Trace Option}, where nothing is pre-allocated and each node
adds \ioam data while expanding the packet as well; ($iii$) the \dfn{Proof of
Transit} (POT) and, ($iv$) the \dfn{Edge-to-Edge} (E2E) Option.
Trace and POT options are both embedded in an IPv6 HopByHop extension header,
meaning they are processed by every node on the path, while E2E option is
embedded in a Destination extension header, which means it is only processed by
the destination node.

\ioam data fields are defined within \ioam namespaces, which are identified by a 
16-bit identifier. They allow devices that are \ioam capable to determine, for 
example, whether \ioam option header(s) need to be processed, which \ioam option 
headers need to be processed/updated in case of multiple \ioam option headers, 
or whether \ioam option header(s) should be removed. \ioam namespaces can be 
used by an operator to distinguish different operational domains. Devices at 
domain boundaries can filter on namespaces to provide for proper \ioam domain 
isolation. They also provide additional context for \ioam data fields, ensuring 
\ioam data is unique, as well as allowing to identify different sets of devices.

In this paper, we provide the first implementation of IPv6 \ioam in the Linux
kernel.  This implementation is important as it provides to operators and IETF
an insight into \ioam practical aspects.   In addition, \ioam can complement
Layer5-7 tracing solutions such as OpenTracing~\cite{opentracing} and
OpenCensus~\cite{opencensus} to create a comprehensive Layer2-7 Tracing
solution.

This paper describes our implementation in Linux kernel
4.12.\footnote{\label{ioam_code}Our implementation is available as open source 
under the terms of the GNU General Public License version 2 (GPL-2.0). See 
\url{https://github.com/IurmanJ/kernel_ipv6_ioam}} We designed our 
implementation to be as efficient as possible, through a new kernel module. This 
paper, also, discusses early results on performance of our \ioam implementation.  
In particular, we show that inserting \ioam headers and data does not
strongly reduce the bandwidth capabilities of an \ioam domain but it comes with
an additional (but reasonable) delay due to \ioam processing.

The remainder of this paper is organized as follows: Sec.~\ref{implem} describes
our open-source implementation of \ioam inside the Linux kernel; Sec.~\ref{eval}
evaluates the performance of our implementations with a few use cases;  finally,
Sec.~\ref{ccl} concludes this paper by summarizing its main achievements.

\section{Linux-Kernel Implementation}\label{implem}
In this section, we carefully describe how we have implemented \ioam within the
Linux kernel 4.12.\footnoteref{ioam_code}  We first cover the \ioam node
registration as well as any \ioam resources allocation (Sec.~\ref{implem.uapi}
and~\ref{implem.node_registration}).  We next explain how packet parsing can be
done efficiently for \texttt{IPv6 Extension Headers}
(Sec.~\ref{implem.eh_parsing}).  We finally focus on how \ioam headers are
inserted and deleted from packets (Sec.~\ref{implem.header_insertion}
$\to$~\ref{implem.header_deletion}). Our implementation is based on \ioam 
drafts~\cite{ioam-opt-draft,ioam-draft}, respectively versions 02 and 05.

\subsection{User Space API}\label{implem.uapi}
The registration process is required for a node to enable \ioam and understand 
it. As long as it is not the case, the node will simply ignore anything that is 
\ioam related. For that purpose, a new \texttt{ioctl} has been created. In order 
to fit in the kernel code, we provide a \texttt{uapi} (\textit{user space API}) 
to facilitate the \texttt{ioctl} call. This also eases the registration process 
from a user point of view. Indeed, as a good practice, everything inside the 
\texttt{uapi} has been documented, since it is what users can see and include in 
their programs. Fig.~\ref{struct.ioam_node} shows the main structure of the 
\ioam \texttt{uapi}.

\begin{figure}[!t]
	\begin{lstlisting}[basicstyle=\scriptsize\ttfamily]
	struct ioam_node{
		__u32 ioam_node_id; // (*\color{comment}{\ioam}*) node identifier
		int if_nb;          //number of (*\color{comment}{\ioam}*) interfaces
		int ns_nb;          //number of known  (*\color{comment}{\ioam}*) namespaces
		int encap_nb;       //number of (*\color{comment}{\ioam}*) insertions
		struct ioam_interface ifs[IOAM_MAX_IF_NB];
		struct ioam_namespace nss[IOAM_MAX_NS_NB];
		struct ioam_encapsulate encaps[IOAM_MAX_NS_NB];
	};
	\end{lstlisting}
\vspace{-0.5cm}	
	\caption{\ioam node registration structure.}
	\label{struct.ioam_node}
\end{figure}

The list of \ioam interfaces (\texttt{ifs}) contains a mapping between a device
and its \ioam identifier chosen by the operator, as well as its \ioam role $\in$
$\{ none, ingress, egress \}$. A \texttt{none} role means that the interface is
an \ioam domain boundary that does not allow incoming \ioam traffic, while
\texttt{ingress} and \texttt{egress} respectively mean that the interface allows
incoming and outgoing \ioam traffic. An \ioam interface can play the role of
both \texttt{ingress} and \texttt{egress} at the same time. The list of \ioam
namespaces (\texttt{nss}) represents all namespaces known by the node and
contains per-namespace data such as its identifier, as well as whether
it should be removed by the node or not. A node will ignore \ioam data from an 
unknown \ioam namespace. As for the list of \ioam insertions (\texttt{encaps}), 
it contains \ioam data that should be inserted in packets, along with their 
\ioam namespace identifier and the egress \ioam interface. The same namespace 
identifier can be inserted multiple times, for different \ioam options, as well 
as several \ioam options in the same namespace.

This structure provides a complete configuration of an \ioam node and allows for
strict filtering due to \ioam interface roles and \ioam egress interfaces when
inserting \ioam headers. The operator has all the possibilities, such as 
configuring overlapping \ioam namespaces, unidirectional, or bidirectional \ioam 
flows.

\subsection{Node Registration}\label{implem.node_registration}
Calling the \texttt{ioctl} triggers several steps. First, the \ioam kernel
module parses and validates data. Then, it builds internal structures based on
received data and every single \ioam resource is allocated per kernel network
namespace. The main goal is to design a solution to allocate any resources 
needed by \ioam, at registration time, in order to keep packet processing as 
fast as possible.

The very first important resource to store is the list of known \ioam 
namespaces, at the node level as any interfaces must be able to access it. Since 
the \ioam namespace lookup occurs each time a packet with \ioam data is 
processed, a kernel hashtable, with the \ioam namespace identifier as the key, 
has been used for that purpose. The trade-off is to find a balanced between 
potential collisions and memory allocation.  We achieve this by forcing the 
hashtable size being four times the maximum number of allowed \ioam namespaces. 
The \ioam node identifier is also stored along with the hashtable, as well as 
pre-allocated paddings, from length 1 to length 7, to speed up \ioam options 
alignment (see Sec.~\ref{implem.header_insertion}).

\begin{figure}[!t]
	\begin{center}
		\includegraphics[width=8.3cm]{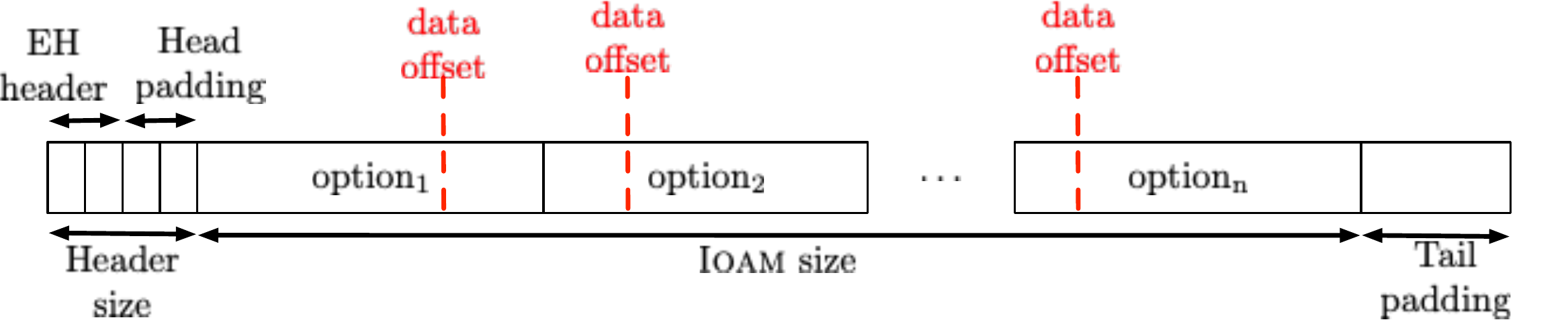}
	\end{center}
\vspace{-0.5cm}	
	\caption{\ioam buffer representing an IPv6 Extension Header.}
	\label{implem.ioam_buffer}
\end{figure}

The other important resource to store is an optimized buffer that represents an
\texttt{Extension Header} with \ioam (either a \texttt{Hop-by-Hop} or
\texttt{Destination} option) to be inserted, built from registration data.
Fig.~\ref{implem.ioam_buffer} illustrates how the buffer is structured and how it
works (more details on this buffer in Sec.~\ref{implem.header_insertion}). It is
allocated twice per \ioam encap interface, one for a \texttt{Hop-by-Hop} option
and another for a \texttt{Destination} option. The \ioam identifier and role are 
also stored there.

\subsection{EH Parsing}\label{implem.eh_parsing}
IPv6 Extension Headers already come with parsing mechanisms inside the kernel,
based on~\cite{rfc8200}. However, for speeding up as much as possible both \ioam 
header insertion and deletion, we need to maintain additional information. This 
is the objective pursued by the \texttt{ioam\_parsed\_eh} structure (see 
Fig.~\ref{struct.ioam_parsed_eh}).  It helps to easily insert and/or remove 
\ioam from packets, while remaining efficient. The usefulness of each field will 
be detailed in Sec.~\ref{implem.header_insertion} 
and~\ref{implem.header_deletion}.

\begin{figure}[!t]
	\begin{lstlisting}[basicstyle=\scriptsize\ttfamily]
	struct ioam_parsed_eh{
		struct ioam_block eh; //Extension Header offset and size  
		struct ioam_block last_pad; //last padding offset and size (if any)
		u8  pad_size; //total padding size		
		u16 decap_size; //total size of (*\color{comment}{\ioam}*) data to remove
		u8  free_idx;
		struct ioam_block decaps[IOAM_MAX_NS_NB];
	};
	\end{lstlisting}
\vspace{-0.5cm}	
	\caption{Structure with additional data for an Extension Header parsing.}
	\label{struct.ioam_parsed_eh}
\end{figure}

The field \texttt{decaps} represents the list of  $\{offset, size\}$ \ioam
blocks to be removed, with \texttt{free\_idx} giving the very next free slot in
the array, as well as the number of \ioam options to be removed. Having a
fixed-size array is a choice we made to avoid allocation during critical
processing. Note that defined boundaries in the \ioam user space API are
realistic and as such are not a problematic limit.

As both the header insertion and deletion need this data, we need to store the
structure, while keeping in mind the main goal of avoiding any
allocation. Initially, we were storing the structure per network devices
(\texttt{struct net\_device}), where it was pre-allocated at the registration.
However, we found out that packet queues were multi-threaded in the kernel,
which prevented this solution from working. Ideally, the structure should follow
its specific packet, just like if it was attached to it, but that would also
mean an allocation during packet processing. We started to look inside the
\texttt{sk\_buff} structure that represents a network packet, and discovered
that a private user data space was available. This field is called the
\texttt{control buffer} (\texttt{cb}) and is said to be free to use for every
layer, where one can put private variables there. But the current layer, IPv6,
already uses this field for storing an internal structure
(\texttt{inet6\_skb\_parm}).  Finally, in order to integrate \ioam as
efficiently as possible within the kernel, we decided to store a pointer to
\texttt{ioam\_parsed\_eh} inside \texttt{cb}. It means less memory to
allocate per packet (4 or 8 octets, depending on the architecture), which is
critical, especially when this additional field aims at being used ``locally''.
A concrete and simplified example of how it works is shown in
Fig.~\ref{code_flow}. Thanks to this technique, we first respect our objective
to not allocate anything during packet processing and, secondly, multi-threaded
packet queues are now supported.

\begin{figure}[!t]
	\begin{lstlisting}[basicstyle=\scriptsize\ttfamily]
	struct ioam_parsed_eh parsed_eh;
	IP6CB(skb)->ioam = &parsed_eh;
	
	if (ipv6_parse_hopopts(skb)){
	  /* decap, if any, based on parsed_eh */
	}
	
	/* encap, if any, based on parsed_eh */
	IP6CB(skb)->ioam = NULL;
	\end{lstlisting}
\vspace{-0.5cm}	
	\caption{Simplified example of IPv6 input for a Hop-by-Hop.}
	\label{code_flow}
\end{figure}

\subsection{Header Insertion}\label{implem.header_insertion}
The \ioam header insertion happens after both the parsing
(Sec.~\ref{implem.eh_parsing}) and the \ioam header deletion
(Sec.~\ref{implem.header_deletion}). The first reason of such a decision is that
we benefit from the parsing. It means that there is no need to do the same job 
again, and we also make sure that data is consistent and well-formed. The second 
reason is that we want to remove data first, if any, before inserting new ones. 
It allows for avoiding a possible allocation if the packet buffer has no free 
room anymore.

Two different cases can be met here. The first one is when there is no
\texttt{Hop-by-Hop} option in the packet. The same logic can be applied to a
\texttt{Destination} option. In that case, we simply insert the entire buffer
shown in Fig.~\ref{implem.ioam_buffer}. The first two octets represent the
\texttt{EH} header, while two next octets are padding. Indeed, \ioam options
require a $4n$ alignment~\cite{ioam-draft,ioam-opt-draft}, which is specified
using the notation $xn+y$, meaning the Option Type must appear at an integer
multiple of $x$ octets from the start of the \texttt{EH} header, plus $y$
octets. Following that, each \ioam option is inserted as well as a tail padding
such that the complete \texttt{EH} length is an integer multiple of 8
octets~\cite{rfc8200}. This is the easiest and fastest case.

The other case is when there is a \texttt{Hop-by-Hop} (or \texttt{Destination})
option in the packet. In that case, we need to insert the \ioam data, but we
first need to compute the head and tail padding required for both alignment and
length. This is where the parsing data is useful, as illustrated in
Fig.~\ref{code_insert}. The extra space to insert is computed based on new head
and tail paddings, who themselves depend on the last padding found in the
\texttt{EH}. We reuse the entire \texttt{EH} minus the tail padding, 
if any, and we append \ioam data with correct head and tail paddings to respect 
both alignment and length.

\begin{figure}[!t]
	\begin{lstlisting}[basicstyle=\scriptsize\ttfamily]
	/* Do we have a padding at the end ? */
	if (lastpad.offset + lastpad.size == eh.offset + eh.size)
		tailpad_size = lastpad.size;
	else
		tailpad_size = 0;
	
	eh_notail_size = eh.size - tailpad_size;
	new_headpad_size = (4 - (eh_notail_size % 4)) % 4;
	new_tailpad_size = ((8 - ((eh_notail_size + ioam_size) % 8)) % 8) - new_headpad_size;
	extra_room = ioam_size - tailpad_size + new_headpad_size + new_tailpad_size;
	\end{lstlisting}
\vspace{-0.5cm}	
	\caption{\ioam header insertion and computations.}
	\label{code_insert}
\end{figure}

The strength of this algorithm is that it does not require any allocation, even
for head and tail paddings, as they are already allocated at registration time. 
However, an implicit allocation can occur if the required extra space is bigger 
than the free space in the packet buffer. A solution would be to increase the
\texttt{needed\_headroom} field of each \ioam encap interface 
(\texttt{struct net\_device}) to allocate a bit more for each packet and as such 
avoid this situation. Again, this is a trade-off between memory and 
performance done by the operator.

\subsection{Data Insertion}\label{implem.data_insertion}
The insertion of \ioam data is only applied by a node if the input interface has 
an \ioam ingress role and if the \ioam namespace is known. The latter is 
required for a lookup inside the hashtable containing all known \ioam namespaces 
for the node. If those conditions are respected, the node inserts its data, 
depending on the asked \ioam mode/option. Some data may be unavailable inside 
the Linux kernel (e.g., queue depth, buffer occupancy).

Note that the data insertion also happens during \ioam header insertion, with
the only difference that, here, the above checks are not applied. Indeed, it is the
node itself that built the buffer. Any additional check in this situation would
be useless and a clear waste of time.

\subsection{Header Deletion}\label{implem.header_deletion}
The \ioam header deletion is straightforward due to data provided by the 
improved parsing. We know the total size of data blocks to be removed as 
well as their respective positions and sizes, as shown in 
Fig.~\ref{struct.ioam_parsed_eh}. One can distinguish two different cases. The
first and fastest one is when the entire \texttt{EH} must be removed. It is 
possible to quickly check it since we know the size of data to be removed, as well as 
the total padding and the \texttt{EH} size. If 
$\texttt{decap\_size} + \texttt{pad\_size} == \texttt{EH.size} - 2$, then the 
entire \texttt{EH} must be removed (minus 2 to not take \texttt{EH} header into 
account). As for the second case, it will omit \ioam blocks that must be removed 
while shifting/copying data, and append tail padding if required.

\begin{figure}[!t]
  \begin{center}
    \subfloat[\textit{before} \ioam deletion]{
      \includegraphics[width=7.5cm]{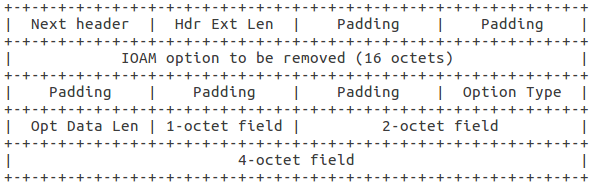}
      \label{implem.deletion.before}
    }
    
    \subfloat[\textit{after} \ioam deletion]{
      \includegraphics[width=7.5cm]{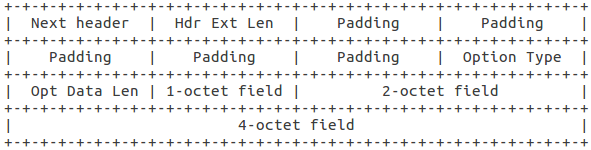}
      \label{implem.deletion.after}
    }
  \end{center}
\vspace{-0.5cm}  
  \caption{\ioam deletion and alignment.}
  \label{implem.deletion}
\end{figure}

Just like the header insertion algorithm, there is no allocation involved here.
Of course, to stay efficient, options are not re-ordered so the arrangement
could not be optimal. Still, it seems that alignment is still respected after
the deletion, for any option that was positioned after a removed \ioam header,
by only removing blocks without taking care of paddings before and after these
blocks. At worst, alignment is respected with unnecessary paddings. An example
is shown in Fig.~\ref{implem.deletion.before} where an \ioam option ($4n$
aligned) and another option ($4n+3$ aligned) are present.
Fig.~\ref{implem.deletion.after} shows the result after the \ioam header
deletion. One can see that the remaining option is still $4n+3$ aligned but
5-octet padding is present, where 1-octet padding would have been enough. Again,
this is not a real problem as it will still work as expected, but this is not
elegant. A possible solution would be to check if the \ioam header to be removed
is in-between paddings and recompute the minimal merged padding size. However,
it would have an impact on memory, as more data would need to be stored, as well
as on performance. Consequently, it may not be worth for it, considering the
usage and goal of \ioam.

\section{Evaluation}\label{eval}
In this section, we make a first attempt at evaluating \ioam performances inside
the Linux kernel.  We first present our testbed (Sec.~\ref{eval.testbed}) and,
next (Sec.~\ref{eval.results}), discuss our results.

\subsection{Testbed}\label{eval.testbed}
\begin{figure}[!t]
	\begin{center}
		\includegraphics[width=8cm]{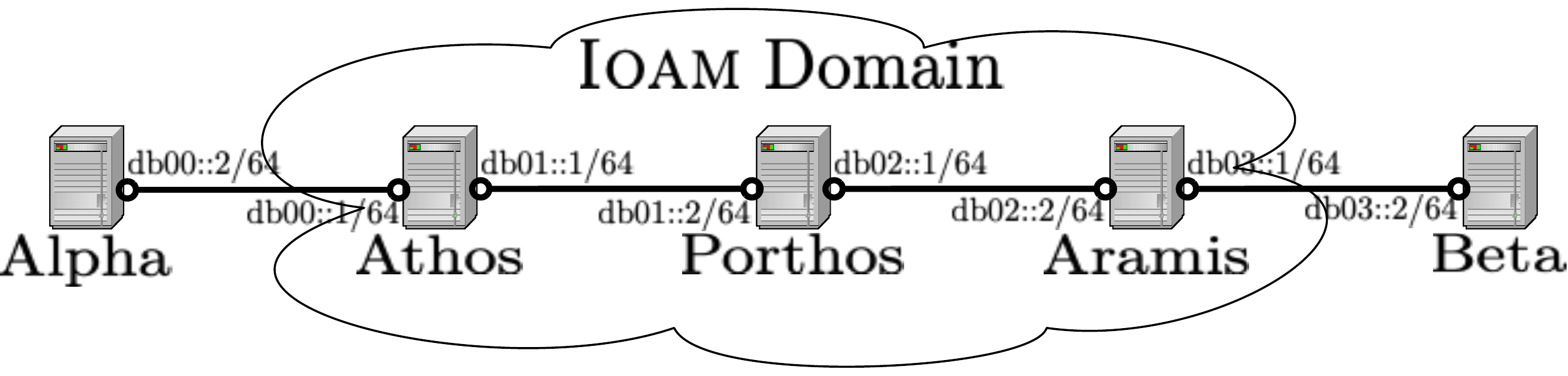}
	\end{center}
	\caption{Topology used for measurements.}
	\label{eval.topology}
\end{figure}

For performing the performance measurements, we setup a testbed on a single
physical machine based on Intel Xeon CPU E5-2683 v4 at 2.10GHz, 16 Cores, 32
Threads, 64GB RAM, running Debian 9.0 with a 4.12 patched kernel to include our
\ioam implementation. Fig.~\ref{eval.topology} shows the five-node topology
running on the machine, where each node is isolated inside a dedicated Linux
container via \texttt{lxc 2.0.7}~\cite{lxc}.

\texttt{Alpha} is the traffic generator (\texttt{TG}), while \texttt{Beta} is
the receiver. Three nodes represent the \ioam domain whose boundaries are
\texttt{Athos} and \texttt{Aramis}. They respectively insert and remove \ioam
headers. \texttt{Porthos} inserts its \ioam data, just as both \texttt{Athos}
and \texttt{Aramis} do too. Containers are linked through veth peers. The
\texttt{TG} runs \texttt{pktgen} 2.75~\cite{pktgen}.  \texttt{pktgen} is
included in the kernel, meaning packets generated are directly handed by the
network driver. \texttt{pktgen} is only able to generate UDP packets.  Packets
are sized to 1,200 bytes.  This is due to the fact that some tests require
insertions of $\approx$ 200 bytes in packets while we want to respect the MTU
(default 1,500 bytes) and measure the same thing all along.

Each experiment lasts 60 seconds and is run 30 times. Each data point in
subsequent plots represents the mean value over those 30 runs.  

\subsection{Results}\label{eval.results}
\begin{figure}[!t]
  \begin{center}
	\includegraphics[width=6.5cm]{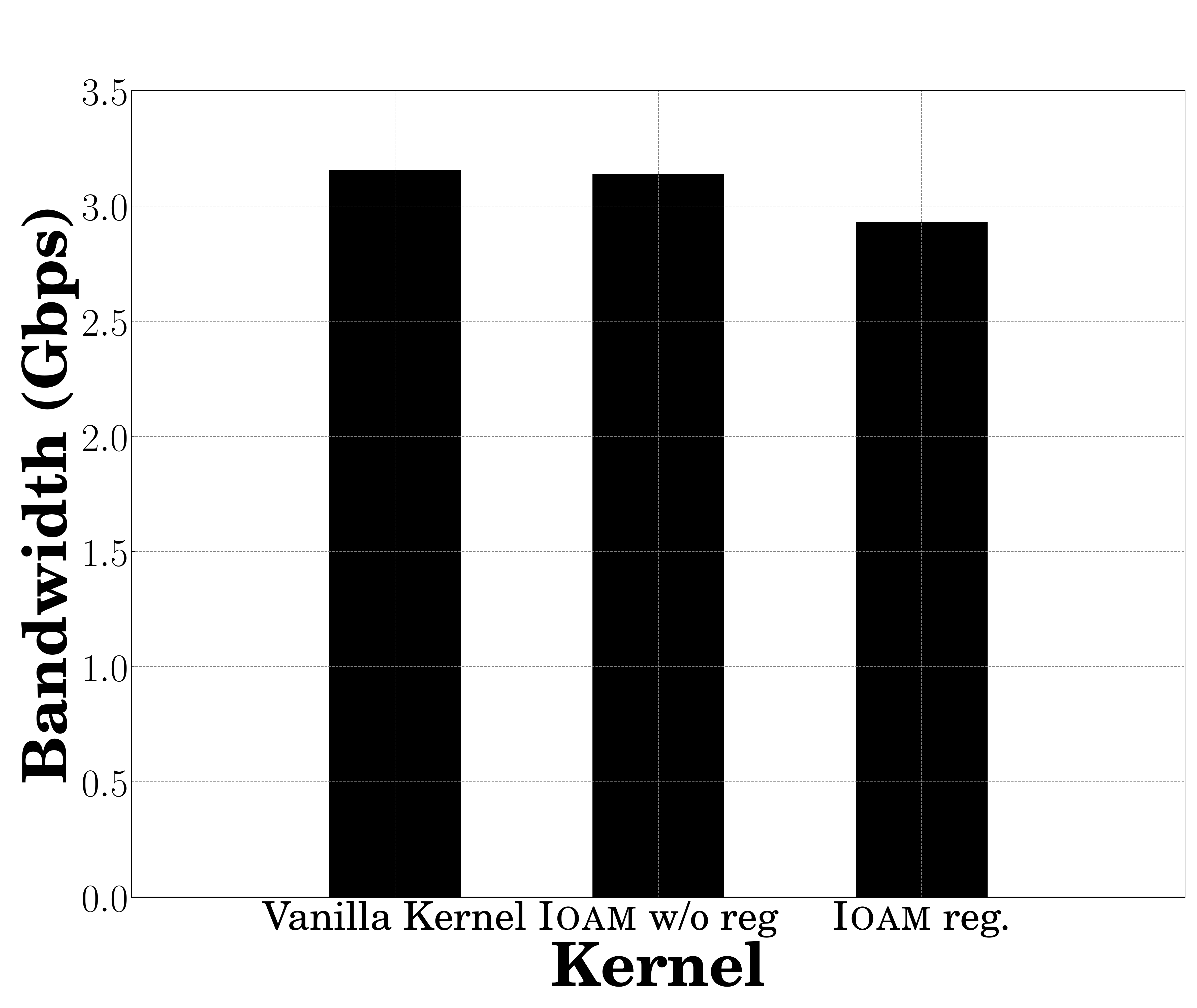}
  \end{center}
\vspace{-0.5cm}  	
  \caption{Bandwidth comparison of \ioam with vanilla kernel.}
  \label{eval.test1}
\end{figure}

We first examine the bandwidth baseline on a vanilla kernel, and compare it to
the same kernel patched with \ioam. We distinguish two cases. Firstly, \ioam
without registration means the \ioam module is enabled in the kernel but none of
the nodes are registered. The goal is to measure the impact of the patch
being totally passive. Secondly, \ioam with registration means that nodes are
registered and, as such, \ioam capable. Fig.~\ref{eval.test1} shows the results
for each measurement. We see a very slight decrease between a vanilla kernel
(3.15 Gbps) and the patched \ioam kernel (3.13 Gbps). Unsurprisingly, the loss
is more visible when nodes are registered (2.93 Gbps), although not that huge
(a drop of 7\%).   In the remainder of this section, for subsequent tests, we
only consider \ioam with registration on traffic that does not contain any 
HopByHop before entering the \ioam domain. 

\begin{figure}[!t]
  \begin{center}
	\includegraphics[width=6.5cm]{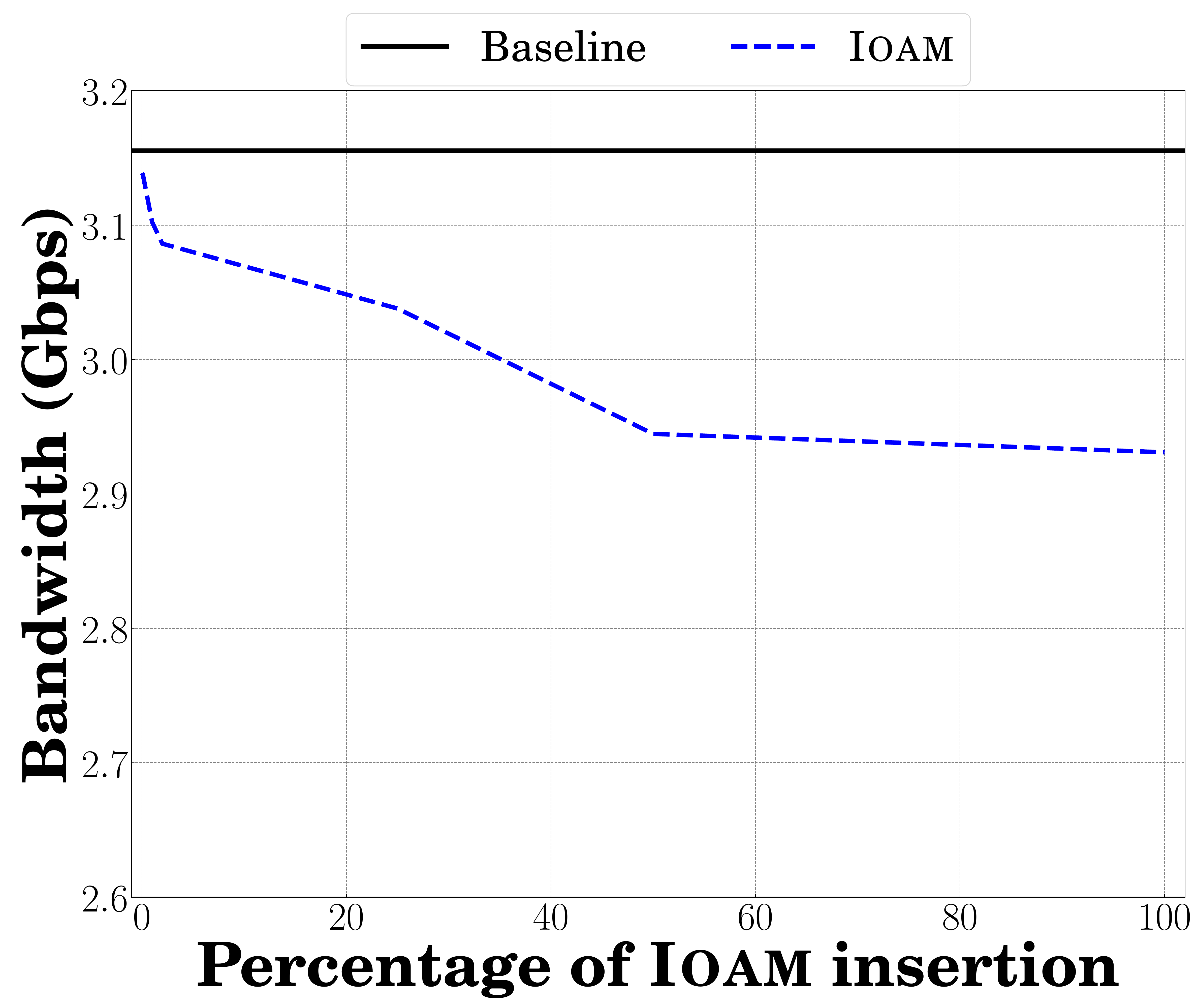}
  \end{center}
\vspace{-0.5cm}  
  \caption{Impact of \ioam data insertion on the overall bandwidth.}
  \label{eval.test2}
\end{figure}

\ioam aims at inserting telemetry data inside IPv6 packet.  We now investigate
the impact of this insertion process, as a fraction of IPv6 packets.  In
particular, we insert \ioam data in IPv6 packets for 0\% (i.e., no \ioam
injection), 0.01\%, 0.1\%, 1\%, 2\%, 10\%, 25\%, 50\%, and 100\% (i.e., \ioam
data is injected in every packet) of the packets traversing the \ioam domain.  We
consider a single namespace containing a single option.  The impact on bandwidth
is plotted on Fig.~\ref{eval.test2}. Without any surprise, ($i$) the bandwidth
reached with 0\% of \ioam injection is consistent with results provided in
Fig.~\ref{eval.test1} and ($ii$) the bandwidth decreases with proportion of
packets involved in \ioam data injection. At worst (all packets traversing the
\ioam domain are involved in \ioam data injection), the bandwidth is dropped by
roughly 7\%. However, this extreme case is unrealistic as we do not expect any
operator inserting telemetry data in every packet traversing its domain (the
data processing load would quickly become unmanageable).  We rather expect \ioam
being applied on a tiny portion of the traffic (or for very specific and limited
in time use case), corresponding so to results on the upper left of
Fig.~\ref{eval.test2}.  In such a case, the impact on the \ioam domain bandwidth
is quite limited (a reduction of 2\% when inserting \ioam in 1\% of the
packets).

\begin{figure}[!t]
  \begin{center}
	\includegraphics[width=6.5cm]{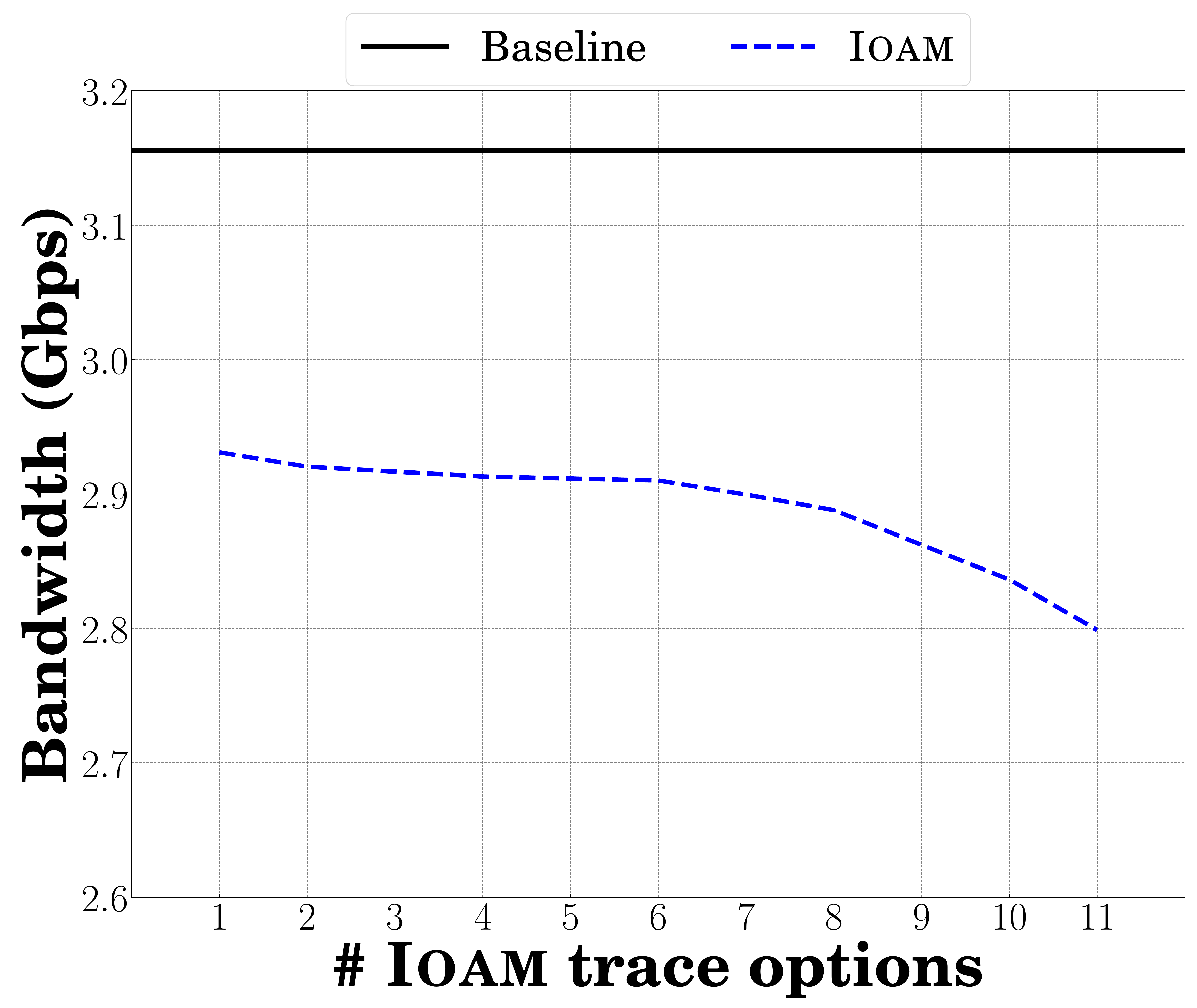}
  \end{center}
\vspace{-0.5cm}  
  \caption{Impact of the number of \ioam Trace options on the overall
  bandwidth.}
  \label{eval.test3}
\end{figure}

We next evaluate the effect of the number of \ioam Trace options on the overall
bandwidth of the \ioam domain. Fig.~\ref{eval.test3} shows the results for 1
$\to$ 11 options. We inject \ioam data in every packets, corresponding so to the
extreme case discussed with Fig.~\ref{eval.test2} (the bandwidth for 1 option is
thus equivalent to the bandwidth achieved with 100\% of packets impacted by
\ioam in Fig.~\ref{eval.test2}). The bandwidth remains globally stable until six
options are inserted and, then, it starts marking a clear decrease.  We believe
that this is due to the fact that inserting more options leads to exceeding the
free room in packets buffer.  It therefore requires a re-allocation of extra
space by the kernel to be able to deal with the entire \ioam data.

\begin{figure}[!t]
  \begin{center}
	\includegraphics[width=6.5cm]{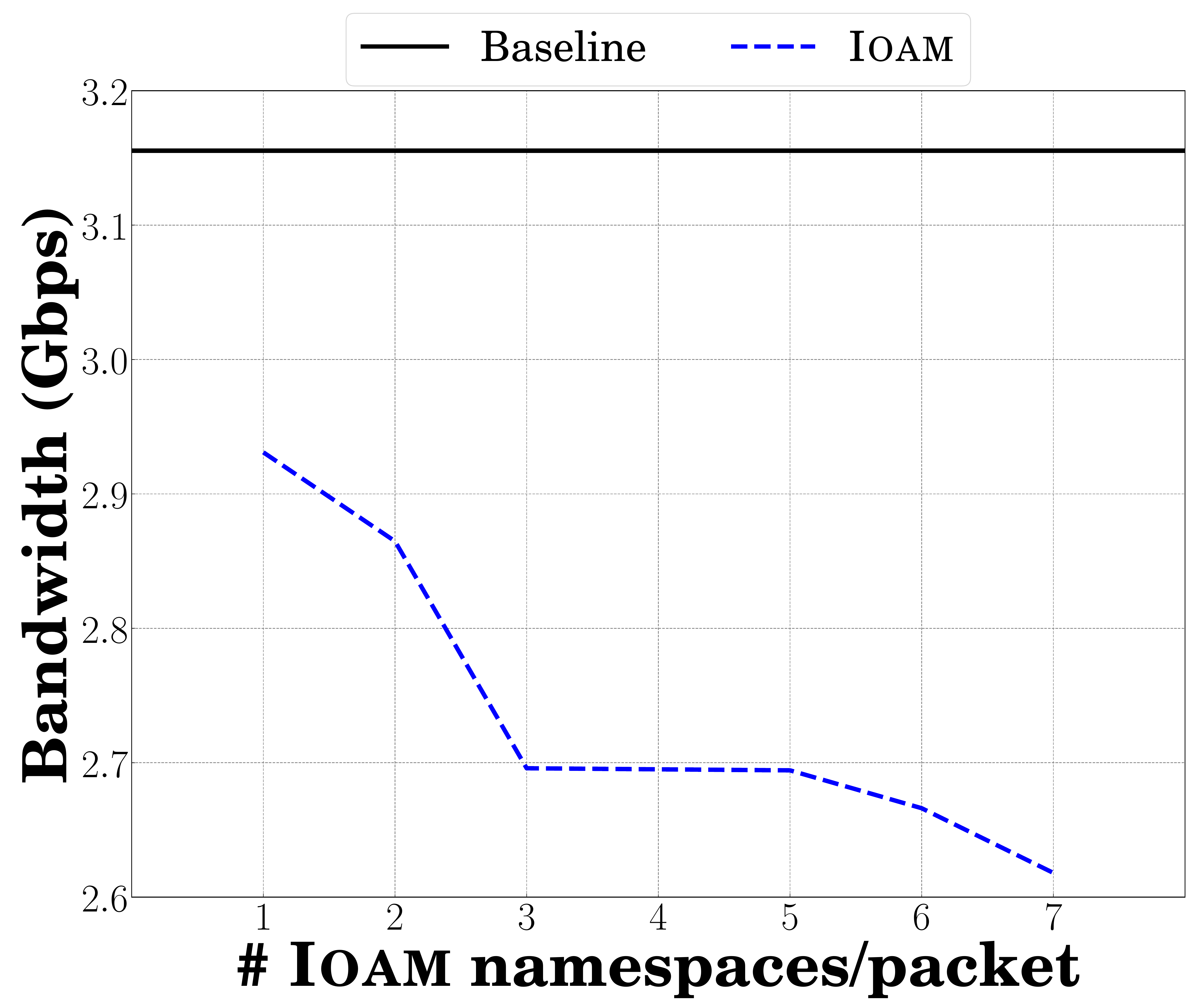}
  \end{center}
\vspace{-0.5cm}	
  \caption{Impact of the number of \ioam namespaces on the overall bandwidth.}
  \label{eval.test4}
\end{figure}

Fig.~\ref{eval.test4} investigates the impact of the number of \ioam namespaces
per packet on the overall bandwidth of the \ioam domain.  For those
measurements, each namespace contains a single \ioam Trace option inserted in
every packet (100\% on Fig.~\ref{eval.test2}), meaning that the bandwidth
achieved for a single namespace corresponds to the one achieved for a single
Trace option in Fig.~\ref{eval.test3}.  On Fig.~\ref{eval.test4}, we can notice
a decrease between two and three \ioam namespaces inserted.  In the fashion of
Trace option, we exceed here the free room in packets buffer with three \ioam
namespaces.  Thus, approximatively, three \ioam namespaces represent the same
amount of bytes than six \ioam Trace options.  The bandwidth remains roughly
stable until five \ioam namespaces, where we observe another drop (attributed
to the fact that there are more and more bytes to insert).

\begin{figure}[!t]
  \begin{center}
	\includegraphics[width=6.5cm]{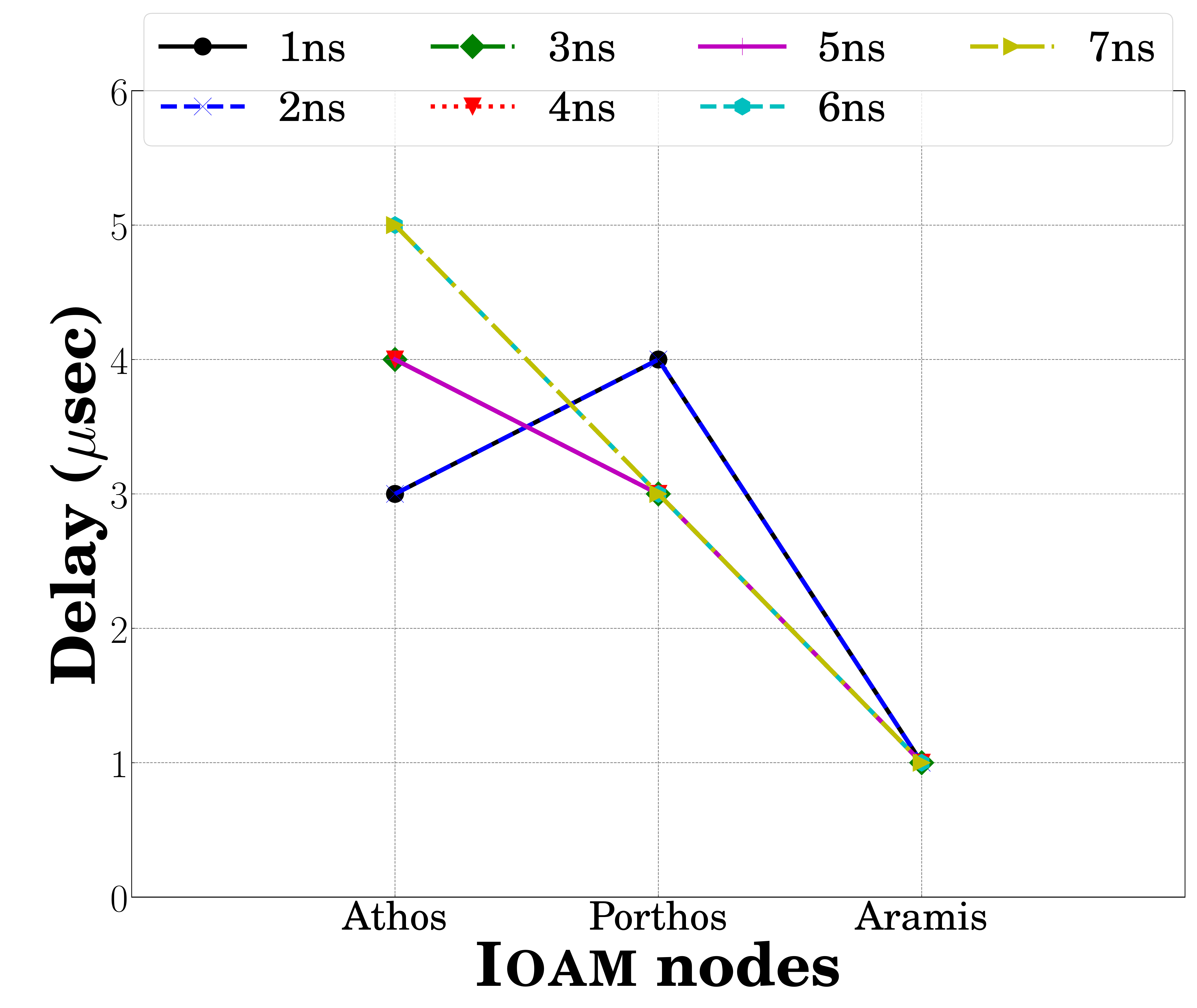}
  \end{center}
\vspace{-0.5cm}  
  \caption{Additional delay due to \ioam operation in each node within the
  \ioam domain.}
  \label{eval.test4delays}
\end{figure}

Finally, we estimate the additional delay due to \ioam processing in each node
within the \ioam domain.  Fig.~\ref{eval.test4delays} reports the minimum delays
when varying the number of namespaces between one and seven.  We first look at
the encap node (i.e, ``Athos'') and see three groups of additional delays: (1,
2) \ioam namespaces, (3, 4, 5) \ioam namespaces, and (6,7) \ioam namespaces.   
The increase in the delay between those groups is explained by the fact that we
are adding more and more bytes in packets, requiring so more processing due to
possible re-allocation.  On the contrary, for the decap node (i.e., ``Aramis''),
the additional processing delay remains the same, whatever the number of
namespaces considered.  This is explained by the fact that the entire IPv6
HopByHop extension header is removed at once, no matter the number of namespaces
inserted inside.  The intermediate node (i.e., ``Porthos'') encounters two
groups of additional processing delay: (1,2,6,7) \ioam namespaces and (3,4,5)
\ioam namespaces.  We ascribe this behavior to some noise in measurements, as we
expect the additional processing delay progressively growing with the number of
namespaces considered (or, at least, be the same).

\section{Conclusion}\label{ccl}
In-situ OAM (\ioam) allows operators, within a pre-defined domain, to insert
telemetry data within data packet, without injecting additional traffic for
measurements.  As such, it has the potential to enhance operations.  \ioam is
still in its infancy as it is still under standardization by the IETF.

This paper reports what is, to the best of our knowledge, the very first
implementation of \ioam for IPv6 in the Linux kernel\footnoteref{ioam_code} as
well as early results on \ioam performance.  We hope that our implementation
supports \ioam standardization, drives adoption of \ioam as well as associated
research. With the availability of \ioam for the Linux Kernel, we can evolve to
a comprehensive tracing solution combining Layer 5-7 tracing (e.g. OpenTracing,
OpenCensus) with \ioam tracing at network level.

\small{
\balance
\bibliographystyle{IEEEtran}
\bibliography{Bibliography}
}

\end{document}